Polymer Entanglement Dynamics: Role of Attractive Interactions

Gary S. Grest*

Sandia National Laboratories, Albuquerque, New Mexico 87185 USA

Abstract

The coupled dynamics of entangled polymers which span a broad time and length scales govern the unique viscoelastic properties of polymers. To follow chain mobility by numerical simulations from the intermediate Rouse and reptation regimes to the late time diffusive regime, highly coarse grained models with purely repulsive interactions between monomers are widely used since they are computationally the most efficient. Here using large scale molecular dynamics simulations the effect of including the attractive interaction between monomers on the dynamics of entangled polymers melts is explored for the first time over wide temperature range. Attractive interactions has little effect on the local packing for all temperatures $T$ and on the chain mobility for $T$ higher than about twice the glass transition $T_g$. These results, across a broad range of molecular weight, show that to study the dynamics of entangled polymers melts the interactions can be treated as pure repulsive, confirming *a posteriori* the validity of previous studies and opening the way to new large scale numerical simulations.

The dynamics of polymers, which is at the core of their unique viscoeselastic behavior, span a wide range of coupled time and length scales. For unentangled chains, topological constraints do not play a dominant role and the chain dynamics can be described theoretically by a Langevin equation with noise and the constraint that monomers are connected to form a chain.[1,2] For longer, entangled chains the dynamics is usually described by the reptation model of Edwards[3] and de Gennes.[4] On length scales larger than the tube diameter, monomers of the chain move predominantly along their own contour. Since this is a one-dimensional diffusion along a random walk path, a chain needs a time $\tau_d \sim N^3/N_e$ to leave the original tube and move diffusively, where $N$ is the chain length and $N_e$ is the entanglement length.

While experiments have been instrumental in testing the macroscopic predictions of the reptation model, there are few experiments which can directly probe the microscopic motion of a polymer chain. Neutron spin echo spectroscopy,[5] which follows the local chain dynamics for hundreds of nanoseconds, is one exception. Computer simulations have played an important role in probing the local dynamics of entangled polymers as they can directly follow the local motion of the chains with the main limitation being the computational resources needed to reach long time. For this reason most studies of entangled polymer melts have used coarse grained models in which the interactions are truncated at very short distances, often with purely repulsive interactions.[6-11] While the chain mobility is expected to be independent of the range of potential at sufficiently high temperature T this cannot persist at lower $T$ where the strength of the attractive interactions become larger than the thermal energy. Previously, molecular dynamics (MD) simulations have been used to model short, unentangled polymer melts with either purely repulsive or attractive interactions as one approaches the glass transiton.[12-19] Common to all computaional studies, the nature of the interactions is at the core of their ability to adequately probe polymeric properties. Therefore accessing the basic assumption that polymer chains can be described solely by repulsive interatcions is one critical aspect of probing polymers computationally. Here the role of the attractive interactions on multiple time scales are probed for the first time in entangled polymer melts.

All the simulations were carried out using the bead-spring model.[6] Each chain contains $N$ beads of mass $m$. All beads interact via the Lennard-Jones (LJ) potential[20]

$$U(r) = 4\varepsilon\left[\left(\frac{\sigma}{r}\right)^{12} - \left(\frac{\sigma}{r}\right)^{6}\right] - 4\varepsilon\left[\left(\frac{\sigma}{rc}\right)^{12} - \left(\frac{\sigma}{rc}\right)^{6}\right] \quad r < r_c$$

$$0 \quad\quad\quad\quad\quad r > r_c$$

where $r$ is the distance between two beads, $\varepsilon$ is the Lennard-Jones unit of energy, and $\sigma$ is the bead diameter. $\tau=\sigma(m/\varepsilon)^{1/2}$ is the characteristic unit of time. The interactions are either purely repulsive with $r_c=2^{1/6}\sigma$ or attractive with $r_c=2.5\sigma$. In the later case there is small discontinuity in the force at the cutoff. Beads along the chain are connected by an additional unbreakable finitely extensible nonlinear elastic (FENE) potential $U_{FENE}(r)=-1/2kR^2_0 \ln[1-(r/R_0)^2]$, with $R_0=1.5\sigma$ and $k=30k_B/\sigma^2$. $k_B$ is the Boltzmann constant. To vary $N_e$, a bond bending potential[21] $U_B(\theta)=k_\theta(1+\cos\theta)$, where $\theta$ is the angle between two consecutive bonds, is included.

All simulations were carried out using the Large-scale Atomic/Molecular Massively Parallel Simulator (LAMMPS) parallel MD code.[22] Melts of 500 chains of $N=500$ with $k_\theta=0$ and 250 chains $N=200$ with $k_\theta=1.5\varepsilon$ were studied. The equations of motion were integrated using a velocity-Verlet algorithm with a time step $\delta t=0.01\tau$ for all simulations except $\delta t=0.012\tau$ for $N=500$ with $r_c=2^{1/6}\sigma$. Temperature was maintained by coupling the system weakly to a Langevin heat bath[23, 24] with a damping constant $\Gamma=0.01\tau^{-1}$ for $N=200$ and $0.5\tau^{-1}$ for $N=500$. The melts were constructed following Auhl et al.[25] with periodic boundary conditions in all three directions. The systems were equilibrated at $T=\varepsilon/k_B$, $P=0.02\varepsilon/\sigma^3$ for $10^4\tau$ with $r_c=2.5\sigma$. The resulting monomer density is $\rho\sigma^3=0.889$ for $N=500$ and $\rho\sigma^3=0.885$ for $N=200$. This density is slightly larger than the monomer density $\rho\sigma^3=0.85$ used in previous simulations[6-9] for the purely repulsive case since for $r_c=2.5\sigma$, ($\rho\sigma^3=0.85$, $T=\varepsilon/k_B$) is in the two phase regime. Melts at lower $T$ were obtained by cooling the system with $r_c=2.5\sigma$ at a rate of $10^{-7}\varepsilon/k_B\tau$ at $P=0.02\varepsilon/\sigma^3$. The length of the runs and densities are listed in Table 1 for all temperatures studied here. Including the attractive interactions increased the cpu time by a factor of 2.1-2.3.

Table I. Results for $N=200$ with $k_\theta=1.5\varepsilon$. $\rho$ is the monomer density, $<R^2>$ is mean squared average end-to-end distance, $p$ is the packing length, $N_e$ is the entanglement length and $\tau_{run}/\tau$ is the length of the run. Results for $T\geq 0.6\,\varepsilon/k_B$ are for $r_c=2.5\sigma$ while those for $T=0.50\,\varepsilon/k_B$ are for $r_c=2^{1/6}\sigma$ except the density which is from the slow cooling with $r_c=2.5\sigma$.

| $k_BT/\varepsilon$ | $\rho\sigma^3$ | $<R^2>/\sigma^2$ | $p/\sigma$ | $N_e$ | $\tau_{run}\times 10^{-7}/\tau$ |
|---|---|---|---|---|---|
| 1.5  | 0.765 | 455 | 0.57 | 50 | 1.6  |
| 1.0  | 0.885 | 532 | 0.42 | 28 | 2.4  |
| 0.75 | 0.952 | 645 | 0.33 | 20 | 3.3  |
| 0.60 | 0.994 | 734 | 0.27 | 16 | 10.3 |
| 0.50 | 1.022 | 923 | 0.21 | 15 | 18.0 |

The primitive path analysis (PPA) algorithm of Everaers et al.[26] was used to identify $N_e$. In the PPA the chain ends are fixed and the intrachain excluded volume interactions are turned off, while retaining the interchain excluded volume interactions. The energy of the system is then minimized by slowly cooling the system to $T=0$. $N_e$ is obtained from $<L_{pp}^2>$, the mean squared contour length of the primitive path, and $<R^2>$, the end-to-end distance of the chain, using the formula,[27] $N_e=(N-1)(<L_{pp}^2>/<R^2>-1)^{-1}$. This equation converges to the asymptotic value of $N_e$ for large $N$ faster then the formula $N_e =(N-1) <L_{pp}>^2/<R^2>$ proposed by Everaers et al.[26] At $T=\varepsilon/k_B$, $N_e\sim 84$ for $k_\theta=0$ and $\sim 28$ for $k_\theta=1.5\varepsilon$, consistent with previous estimates.[27,28] The number of entanglements per chain Z $=N/N_e \sim$ 6 for $N=500$ and $Z \sim$ 7.2 for $N=200$ at $T=\varepsilon/k_B$ due to the difference in $k_\theta$.

Weeks, Chandler and Andersen[29] first showed that the repulsive force dominates the structure of simple liquids and at high density the attractive component of the force has little influence. For a LJ fluid, they showed that including the attractive interactions effects only the low wave vector component of the liquid structure factor and is negligible for $\rho\sigma^3>0.65$. As connecting LJ monomers to form a bead-spring chain increases the equilibrium density compared to a fluid of monomers, it is not surprising that the attractive interactions have hardly any affects the local packing for $T=\varepsilon/k_B$ as observed in the intermolecular pair correlation function g(r) in the inset of Figure 1. Including the attractive tail of the LJ potential increases slightly the height of the first peak and depth of the first minimum compared to purely repulsive potential. $<R^2>$ is independent of $r_c$.

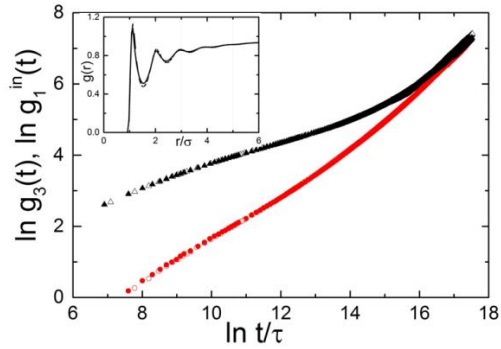

Figure 1. Mean squared displacement of the center of mass $g_3(t)$ (red circles) and inner 10 beads $g_1^{in}(t)$ (black triangles) for chains of length $N=500$ with $k_\theta=0$ for $r_c=2^{1/6}\sigma$ (open symbols) and $r_c=2.5\sigma$ (closed symbols) at $T=\varepsilon/k_B$ and $\rho\sigma^3=0.889$. Pair correlation function g(r) versus r is shown in the inset for $r_c=2^{1/6}\sigma$ (solid) and $2.5\sigma$ (dashed).

At $T$ significantly above $T_g$, the similarity in the monomer packing is expected to transpose to dynamical properties. The mean-squared displacement (MSD) of the center of mass (cm) $g_3(t)=\langle(r_{cm}(t)-r_{cm}(0))^2\rangle$ and the center 10 beads of the chain $g_1^{in}(t)=\langle(r_i(t)-r_i(0))^2\rangle$ are shown in Figure 1 for *N=500* for $r_c=2^{1/6}\sigma$ and $2.5\sigma$ at $T=\varepsilon/k_B$. These results show that from earliest times to the late time diffusive regime the cutoff has essentially almost no effect on the MSD. The results for $g_1^{in}(t)$ show a crossover from the early time $t^{1/2}$ Rouse regime to $t^{1/4}$ scaling predicted by the tube model[2] at an entanglement time $\tau_e\sim10^4\tau$. The two curves converge at a diffusive time $\tau_d\sim1.1\times10^7\tau$, defined by $g_3(\tau_d)=3\langle R^2_g\rangle$. The diffusion constant for the purely repulsive case $D_{rep}=6.4\times10^{-6}\tau/\sigma^2$ is slightly larger than when attractive interactions are included, $D_{att}=5.7\times10^{-6}\tau/\sigma^2$. These results are consistent with Kalathi et al.[30] who showed that for this same system the cutoff has little effect on the Rouse modes of the chain and $N_e$. These results prove that the implicit assumption made in previous equilibrium simulations[6-9] of entangled polymer melts that truncating the potential has no effect on the underlying phenomena, specifically the role of entanglements on the intermediate dynamics. A separate study is needed to determine if this also holds at high shear rates.[31]

To correlate the model to experiment, $T_g$ is determined as a reference point. The two systems were cooled slowly at $P=0.02\varepsilon/\sigma^3$ from $T=\varepsilon/k_B$ to $0.2\varepsilon/k_B$. As shown in Figure 2, above $T_g$ the monomer density and coefficient of thermal expansion do not depend on $k_\theta$. From the break in the slope of the density at high and low $T$, $T_g\sim0.43\varepsilon/k_B$ for the fully flexible model and $\sim0.48\varepsilon/k_B$ for $k_\theta=1.5\varepsilon$. No detectable difference in $T_g$ is observed for cooling at a rate of $10^{-6}\varepsilon/k_B\tau$ compared to $10^{-7}\varepsilon/k_B\tau$. The increase in $T_g$ with increasing $k_\theta$ agrees with Schnell et al.[17] for $N\leq32$. Thus the previous studies[6-9] for this model at temperatures $T=\varepsilon/k_B>2T_g$. This is comparable to experimental studies of many common polymers such as polyethylene,[32,33] polypropylene[34] and polydimethysiloxane[35] which are often studied for $T\sim2\text{-}3T_g$.

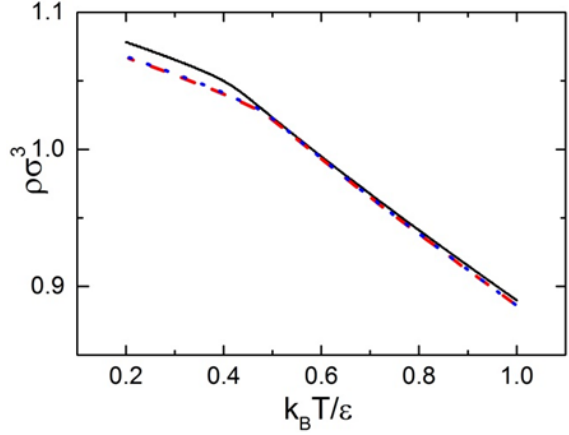

Figure 2. Monomer density $\rho$ as a function of temperature $T$ during cooling at $P=0.02\varepsilon/\sigma^3$ at a cooling rate $10^{-6}\ \varepsilon/k_B\tau$ for $N=500$ with $k_\theta=0$ (black solid) and $10^{-6}\ \varepsilon/k_B\tau$ (red dashed) and $10^{-7}\ \varepsilon/k_B\tau$ (blue dotted) for $N=200$ with $k_\theta=1.5\varepsilon$.

To explore the dependence of the chain mobility on temperature, melts at $T=0.5$, $0.6$, and $0.75\varepsilon/k_B$ obtained at the slowest cooing rate for $N=200$ were simulated at constant volume with purely repulsive and attractive interactions for the same density $\rho$. $N=200$ with $k_\theta=1.5\varepsilon$ was used for this study instead of the fully flexible chains with $N=500$, since the time to reach the diffusive regime is lower even though $Z$ is larger. Results comparing the MSD of the inner 10 monomers of the chain $g_1^{in}(t)$ for 4 temperatures from 0.6 to $1.0\varepsilon/k_B$ for the two cutoff are shown in Figure 3. For all $T$, the mobility decreases with decreasing $T$, however the mobility is always larger for the purely repulsive potential than for the attractive potential with the difference becoming more pronounced as $T$ approaches $T_g$. For $T \leq 0.6\varepsilon/k_B$, $g_1^{in}(t)$ for the two cutoff are well separated over the entire time domain. The differences being due to the inclusion of the attractive interactions on the local monomeric friction coefficient since $\rho$ at each $T$ is the same for both cutoffs.

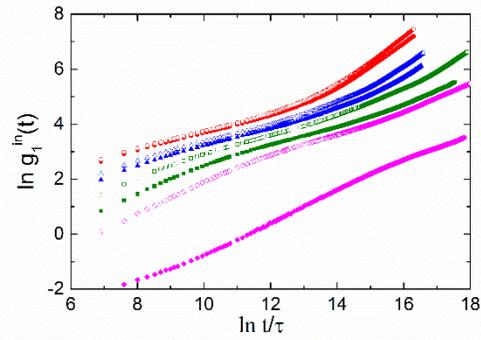

Figure 3. Mean square displacement averaged over the inner 10 beads for $N=200$ with $k_\theta=1.5\varepsilon$ for $r_c=2^{1/6}\sigma$ (open symbols) and $r_c=2.5\sigma$ (closed symbols) for $T=1.0$ (red circles), 0.75 (blue squares), 0.60 (green triangles) and $0.50\varepsilon/k_B$ (magenta diamonds).

The diffusion constant versus $1/T$ are shown in Figure 4. For $T=0.5\varepsilon/k_B$ the diffusion is too small to determine $D_{att}$. The diffusion is approximately Arrhenius for the purely repulsive case and for high temperatures for the attractive case. Derivations from Arrhenius behavior occur only for the later for $T<0.7\ \varepsilon/k_B$. Note that the effect of the cutoff is somewhat larger with $k_\theta>0$ compared to the fully flexible case as seen by comparing $g_1^{in}(t)$ in Figures 1 and 3. For $T=\varepsilon/k_B$, $D_{att}/D_{rep}\sim 0.78$ for $N=200$, $k_\theta=1.5\varepsilon$ compared to $D_{att}/D_{rep}\sim 0.89$ for $N=500$, $k_\theta=0$.

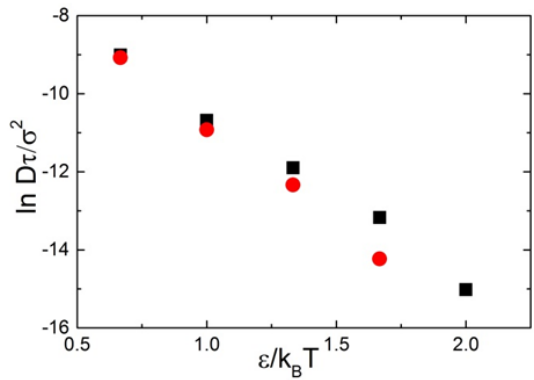

Figure 4. Diffusion constant $D$ versus inverse temperature $1/T$ for $N=200$ with $r_c=2^{1/6}\sigma$ (black squares) and $r_c=2.5\sigma$ (red circles).

The dependence of the packing length $p=N/<R^2>\rho$,[36] and $N_e$ on $T$ are listed in Table 1 for $N=200$. Due to the non-zero bending term, there is a significant increase in $<R^2>$ as T decreases, more than doubling from $455\sigma^2$ at $T=1.5\varepsilon/k_B$ to $923\sigma^2$ at $T=0.5\varepsilon/k_B$. For the fully flexible model the change in $<R^2>$ is only ~5% for the same range of T.[12] This increase in $<R^2>$ results in a decrease in p from $p/\sigma=0.57$ at $T=1.5\varepsilon/k_B$ to 0.21 at $T=0.5\varepsilon/k_B$ even though the monomer density increases as $T$ decreases. This decrease in p with decreasing temperature has been observed experimentally for many polymers.[36] Thus the number of individual chains present in a given volume of the melt decreases with decreasing temperature due to the strong increase $<R^2>$ as $T$ decreases for $k_\theta=1.5\varepsilon$. Over for the same $T$ range, $N_e$ decreases from ~50 to 15.

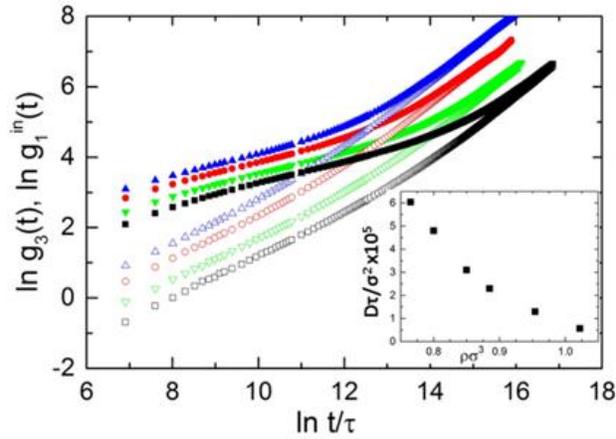

Figure 5. Mean squared displacement of the center of mass $g_3(t)$ (open) and inner 10 beads $g_1^{in}(t)$ (closed) for chains of length $N=200$ with $k_\theta=1.5\varepsilon$ for $\rho\sigma^3=0.765$ (blue up triangles), 0.85 (red circles), 0.95 (green down triangles) and 1.022 (black squares) at $T=\varepsilon/k_B$ for purely repulsive LJ potential with $r_c=2^{1/6}\sigma$. Diffusion constant $D$ versus monomer density $\rho$ is shown in inset.

The density of a polymeric melt is one critical property that strongly impacts its properties. Most previous studies using this model have been largely at a single monomer density $\rho\sigma^3=0.85$. While increasing $\rho$ decreases $N_e$ which is favorable computationally for exploring the reptation regime, it also decreases $D_{rep}$ and increases $\tau_d$ which are unfavorable. This density was chosen since it is relatively high and comparable to the triple point density for an attractive Lennard-Jones monomer fluid.[6] To examine the effect of $\rho$ on the dynamics, simulations at $T=\varepsilon/k_B$ with $r_c=2^{1/6}\sigma$ were run for several densities from $\rho\sigma^3=0.765$ to 1.022 for $N=200$. While varying $\rho$ over this range has only a small effect on the chain statics as $<R^2>$ decreases from $537\sigma^2$ to $524\sigma^2$ and $p$ decreases from $0.49\sigma$ to $0.37\sigma$ as $\rho\sigma^3$ increase from 0.765 to 1.022, it has a large effect on the dynamics as shown in Figure 5. Increasing $\rho$ expands the intermediate reptation scaling regime as $\tau_e$ increases by about a factor of 2 while $\tau_d$ increases by about a factor of 10. $D_{rep}$ decreases by a factor ~12 over this range of density as shown in the inset of Figure 5, while $N_e$ decreases from ~31 for $\rho\sigma^3=0.765$ to ~23 for $\rho\sigma^3=1.022$. Increasing $\rho$ from the canonical $\rho\sigma^3=0.85$ to 1.022 increases $Z$~7.2 to 8.7 for $N=200$ while $\tau_d$ increases from $1.3\times10^6\tau$ to $7.6\times10^6\tau$, roughly a factor of 6. The tube model[2] predicts that $\tau_d\sim N^3/N_e$, which only accounts for 20% of the increase in $\tau_d$. Most of the increase is due to an increase in the local monomeric friction coefficient as a result of the increase in local monomer packing. This can be seen from the differnce in the MSD at $1000\tau$. Computationally this means that to increase $Z$, it is more efficient to increase $N$, then the density. From a practical standpoint of exploring reptation dynamics, the original choice[6] of $\rho\sigma^3=0.85$ is close to optimum.

In conclusion, large scale MD simulations of entangled polymer melts showed that including the attractive component of the non-bonded interaction between monomers has little effect on the local packing and single chain statics for all $T$ when compared to the purely repulsive interaction at the same density. For $T$ larger than about $2T_g$ there is also no significant effect on the chain mobility. However for lower $T$, the attractive interactions play a significant role, reducing the chain mobility compared to the purely repulsive case. Many flexible polymers such as polyethylene and polydimethylsiloxane which have $T_g$ far below room temperature are often studied experimentally for $T$ ~2-3$T_g$. These results support the implicit assumption made in most previous numerical studies of entangled melts that using purely repulsive interactions far above $T_g$ can be used to separate the effect of entanglements on the dynamics from that of the glass transition. Encompassing extended temperature and density ranges this study has shown that polymer melt dynamics can de described by repulsive interactions over extended time and length scales and validates the long standing working assumption of many previous studies.

ACKNOWLEDGMENTS

I thank Kurt Kremer, Dvora Perahia and Mark J. Stevens for helpful discussions. This work was supported by the Sandia Laboratory Directed Research and Development Program. This work was performed at the Center for Integrated Nanotechnologies, a U.S. Department of Energy, Office of Basic Energy Sciences user facility. Sandia National Laboratories is a multiprogram laboratory managed and operated by Sandia Corporation, a Lockheed-Martin Company, for the U.S. Department of Energy under Contract No. DEAC04- 94AL85000.

*E-mail: gsgrest@sandia.gov.